\documentstyle[prl,preprint,aps,epsf]{revtex}
\begin{document}
%\preprint{KNTP-99-02}
\bibliographystyle{plain}
\title{Logarithmic Correction to Newton Potential in Randall-Sundrum Scenario}
\author{D. K. Park$^{1}$\footnote{e-mail:
dkpark@hep.kyungnam.ac.kr}
and S. Tamaryan$^{2}$\footnote{e-mail: sayat@moon.yerphi.am}} 
%H. J. W. M\"{u}ller-Kirsten$^{3}$\footnote{e-mail:
%mueller1@physik.uni-kl.de}
%}
\address{
1.Department of Physics, Kyungnam University, Masan, 631-701, Korea\\
2.Theory Department, Yerevan Physics Institute, Yerevan-36, 375036, Armenia}
%\\
%3.Department of  Physics, University of Kaiserslautern, 
%  67653 Kaiserslautern, Germany 
%}
\date{\today}
\maketitle

\begin{abstract}
Using a fixed-energy amplitude in Randall-Sundrum single brane scenario,
we compute the Newton potential on the brane. It is shown that the 
correction terms to the Newton potential involve a logarithmic factor. 
Especially, when the distance
between two point masses are very small compared to $AdS$ radius, the 
contribution of KK spectrum becomes dominant compared to the usual
inversely square law. This fact may be used to prove the existence of an
extra dimension experimentally.
\end{abstract}
\maketitle
% \tableofcontents
% \listoffigures
\newpage
% %
%\section{Introduction}
Recently, much attention is paid to the Randall-Sundrum(RS) brane-world
scenario\cite{rs99-1,rs99-2}. Especially, RS solved the linearized 
gravitational fluctuation equation and used the result to compute the 
Newton potential on the $3$-brane
\begin{equation}
\label{newton1}
V_{N,RS} = G_N \frac{m_1 m_2}{r} 
\left( 1 + \frac{R^2}{r^2} \right)
\end{equation}
where $R$ is a radius of $AdS_5$. The first term in r.h.s. of 
Eq.(\ref{newton1}) is an usual gravitational potential contributed from 
zero mode of the gravitational fluctuation equation. The next term is a
correction to the Newton potential contributed from the continuum
KK spectrum. 

The slightly different correction to the Newton potential is again
derived in Ref.\cite{garr00,gidd00} with consideration of the bending
effect of the $3$-brane and in Ref.\cite{duff00} by computing the 
one-loop corrections to the gravitational propagator. In both cases 
the final Newton potential is different from Eq.(\ref{newton1}) by a 
constant factor as follows;
\begin{equation}
\label{newton2}
V_{N,GT} = G_N \frac{m_1 m_2}{r}
\left( 1 + \frac{2}{3} \frac{R^2}{r^2} \right).
\end{equation}

In this short letter we argue that 
the correction to the Newton potential on the brane involves a logarithmic
factor. Especially, for the short-range gravity this factor is dominant
compared to the usual inversely square law. This means the contribution of 
KK continuum becomes significant in this limit.

Our starting point is a fixed-energy amplitude derived in 
Ref.\cite{park01,park02}:
\begin{equation}
\label{fixedE}
\hat{G}[R, R: E] = R (\Delta_0 + \Delta_{KK})
\end{equation}
where
\begin{eqnarray}
\label{zero-con}
\Delta_0&=&\frac{2}{m^2 R^2}        \\   \nonumber
\Delta_{KK}&=& \frac{1}{mR}
               \frac{K_0(mR)}{K_1(mR)}
\end{eqnarray}
and $K_{\mu}(z)$ is an usual modified Bessel function, and $m$ is a mass
parameter introduced from the linearized fluctuation equation. It is 
worthwhile explaining briefly how Eq.(\ref{fixedE}) is derived at this
stage.
Firstly, we treated the gravitational fluctuation equation as a 
pure quantum-mechanical Schr\"{o}dinger equation
and transformed the singular nature of the $3$-brane into the singular
behavior of Hamiltonian operator in quantum mechanics. 
Assuming that the extra dimension is a 
single copy of $AdS_5$, {\it i.e.} $AdS$/CFT setting, we have shown 
in Ref.\cite{park01,park02}
that 
the fixed-energy amplitude, which is a Laplace transform of the gravitational
propagator in Euclidean time, is crucially dependent on the boundary
condition(BC) at the location of brane, which is parametrized by a single 
real parameter $\xi$. Setting $\xi = 1/2$, which
means the Dirichlet and Neumann BCs are included with equal weight, one can
derive Eq.(\ref{fixedE}). 

In this letter we would like to derive a Newton potential localized on the
$3$-brane from Eq.(\ref{fixedE}). From Ref.\cite{csa00} the Newton potential
on the brane is approximately\footnote{When Eq.(\ref{fixedE}) is derived in 
Ref.\cite{park01,park02}, we have used a translation of the extra dimension
as $z = y + R$. Thus, the location of the 3-brane in this new coordinate
is $z = R$. That is why we consider $\psi_m(R)$ in Eq.(\ref{newton3}).}
\begin{equation}
\label{newton3}
V(r) \sim G_N \frac{m_1 m_2}{r}
\left[ 1 + \frac{2}{3} \int_{m_0}^{\infty} dm R e^{-m r} |\psi_m(R)|^2
                                                            \right],
\end{equation}
where the first term of Eq.(\ref{newton3}) is contributed from $\Delta_0$
in Eq.(\ref{zero-con}) and next term is contributed from the continuum
KK states. Thus, the problem reduces to compute $\psi_m(R)$ from 
$R \Delta_{KK}$. 

Since, in general, the fixed energy amplitude is represented
as
\begin{equation}
\label{gen-rep}
\hat{G}[x, y; E] = \sum_n 
\frac{\phi_n(x) \phi_n^{\ast}(y)}{\frac{m^2}{2} + E_n}
\end{equation}
for the discrete states, one can derive an eigenstate $\phi_n(x)$ by 
computing the residue of $\hat{G}[x, y; E]$. For the continuum states, however,
we need an appropriate integral representation. For example, let us consider
the simple free particle case whose fixed-energy amplitude is 
\begin{equation}
\label{freeE}
\hat{G}_F[x, y; E] = \frac{e^{-m |x - y|}}{m}
\end{equation}
where $E = m^2 / 2$.
Hence, the problem is how to convert Eq.(\ref{freeE}) into the continuum 
version of Eq.(\ref{gen-rep}). Thus we want to find an integral representation
of $\hat{G}_F[x, y; E]$ as follows;
\begin{equation}
\label{int-rep1}
\hat{G}_F[x, y; E] = \int dk 
\frac{\phi_k(x) \phi_k^{\ast}(y)}{\frac{m^2}{2} + \frac{k^2}{2}}.
\end{equation}
 
After assuming that $\phi_k(x) = {\cal N} e^{i k x}$, one can perform the
integration (\ref{int-rep1}) in the complex $k$-plane and finally find
$|\phi_m(0)|^2 = 1 / 2 \pi$ which makes the nomalization constant
${\cal N}$ of the continuum state to be ${\cal N} = 1 / \sqrt{2 \pi}$. 
This nomalization factor corresponds to a classical picture in which a 
particle starts from a particular point with all momenta equally
likely\cite{fey65}. 
Note that the normalization constant ${\cal N}$ is independent of 
the energy parameter $m$ in the free particle case. 
However, it is not a generic property of quantum 
mechanics, {\it i.e.} the normalization is dependent on $m$ for the general 
case as will be shown in RS2 case.

In the real calculation for the correction to the Newton potential we do not
need an explicit form of the continuum wave function. 
In fact, what we need is a value of the continuum wave function at
a particular point. Thus, for the computation of the Newton potential
it is sufficient to know the fixed-energy amplitude at the location of the
3-brane.

Applying the same procudure to $R \Delta_{KK}$ it is straightforward
to derive $\psi_m(R)$ as follows
\begin{equation}
\label{psiR}
|\psi_m(R)|^2 = \frac{1}{2 \pi} \bigg| \frac{H_0^{(1)} (mR)}
                                            {H_1^{(1)} (mR)} \bigg |
\end{equation}
where $H_{\nu}^{(1)}(z)$ is an usual Hankel function. 
As commented earlier, $|\psi_m(R)|^2$ is dependent on the parameter $m$.
When deriving 
Eq.(\ref{psiR}) we have used a relation of the Hankel function to the
modified Bessel function
\begin{equation}
\label{relation1}
K_{\nu}(z) = \frac{i \pi}{2} e^{\frac{1}{2} \nu \pi i} H_{\nu}^{(1)} (i z).
\end{equation}
Inserting Eq.(\ref{psiR}) into Eq.(\ref{newton3}) with $m_0 = 0$ because of 
no mass gap in RS2 picture, the Newton potential becomes
\begin{eqnarray}
\label{newton4}
V_N&=& G_N \frac{m_1 m_2}{r} \left( 1 + \Delta V \right)  
                                                      \\   \nonumber
\Delta V&=&\frac{1}{3 \pi} \int_0^{\infty} du
\bigg | \frac{H_0^{(1)}(u)}{H_1^{(1)}(u)} \bigg | e^{-\frac{r}{R} u}
\end{eqnarray}
where $u = m R$. 
It seems to be impossible to get an analytical expression for $\Delta V$.
Since, however, the light mass in KK spectrum should make a dominant
contribution
to $\Delta V$, we can use the asymptotic formula
\begin{eqnarray}
\label{6}
\lim_{u \rightarrow 0} H_1^{(1)}(u) \sim -\frac{2 i}{\pi} \frac{1}{u}
                                                            \\  \nonumber
\lim_{u \rightarrow 0} H_0^{(1)}(u) \sim \frac{2 i}{\pi} \ln u.
\end{eqnarray}
Then, the modification of the Newton potential $\Delta V$ reduces to 
approximately
\begin{equation}
\label{7}
\Delta V \sim \frac{1}{3\pi} 
\int_0^{\infty} du u |\ln u| e^{-\frac{r}{R} u}.
\end{equation}
Using the integral formula
\begin{eqnarray}
\label{8}
\int_0^{\infty} x^{\nu -1} \ln x e^{-\mu x} dx&=& \frac{1}{\mu^{\nu}}
\Gamma(\nu) \left[\psi(\nu) - \ln \mu \right]
                                             \\   \nonumber
\int_0^1 x \ln x e^{-\mu x} dx&=& \frac{1}{\mu^2}
\left[ 1 - e^{-\mu} - \gamma + Ei(-\mu) - \ln \mu \right]
\end{eqnarray}
where $\psi(z)$, $\gamma$, and $Ei(z)$ are Digamma function, Euler's constant,
and Exponential-Integral function respectively, the r.h.s. of Eq.(\ref{7}) can
be analytically computed, which yields
\begin{equation}
\label{modpoten1}
\Delta V \sim f\left(\frac{r}{R}\right) \frac{R^2}{r^2}
\end{equation}
where
\begin{equation}
\label{10}
f\left(\frac{r}{R}\right) = \frac{2}{3\pi}
\left[-1 + \gamma + \ln \frac{r}{R} + 2 e^{-\frac{r}{R}} 
      -2 Ei\left(-\frac{r}{R}\right) \right].
\end{equation}

Fig. 1 shows $r/R$-dependence of $f(r / R)$ which indicates that the 
short-range and long-range behaviors of gravitational potential are
completely different from each other. When two point masses on the brane 
are separated with a great distance, {\it i.e.} $r/R \rightarrow \infty$,
Eq.(\ref{10}) indicates 
\begin{equation}
\label{17}
\Delta V \sim \frac{2}{3\pi} \frac{R^2}{r^2} \ln \frac{r}{R}.
\end{equation}
Thus as stressed earlier the modification of the Newton potential involves
a logarithmic factor. Although this logarithmic factor is extremely small
due to the power term, it can be measured by an appropriate experimental 
setting. When the distance between the point masses are very small,
{\it i.e.} $r /R \rightarrow 0$, Eq.(\ref{10}) indicates 
\begin{equation}
\label{17}
\Delta V \sim -\frac{2}{3\pi} \frac{R^2}{r^2} \ln \frac{r}{R}
\end{equation}
which yields very strong attractive force compared to the usual inversely 
square law. This means the contribution of the KK spectrum is dominant for
the short range gravitational potential.

The checking of the logarithmic correction to the Newton potential, especially
the short-range behavior of it, by experiment may give a strong evidence 
for the existence of KK spectrum and as a result, the existence of the extra
dimension.

{\bf Acknowledgement:} 
This work was supported by the Korea Research Foundation
Grant (KRF-2002-015-CP0063).

\newpage

\centerline{\bf Figure Captions}

\vspace{0.9cm}

\noindent
{\bf Figure 1}

Plot of $f(r/R)$. This figure indicates that the short-range and long-bahaviors
of the gravitational potential are completely different from each other. 
``RS'' and ``TG'' in this figure stand for Randall-Sundrum and Garriga-Tanaka.
Thus, these points represent Eq.(\ref{newton1}) and Eq.(\ref{newton2})
respectively.

\newpage
\epsfysize=20cm \epsfbox{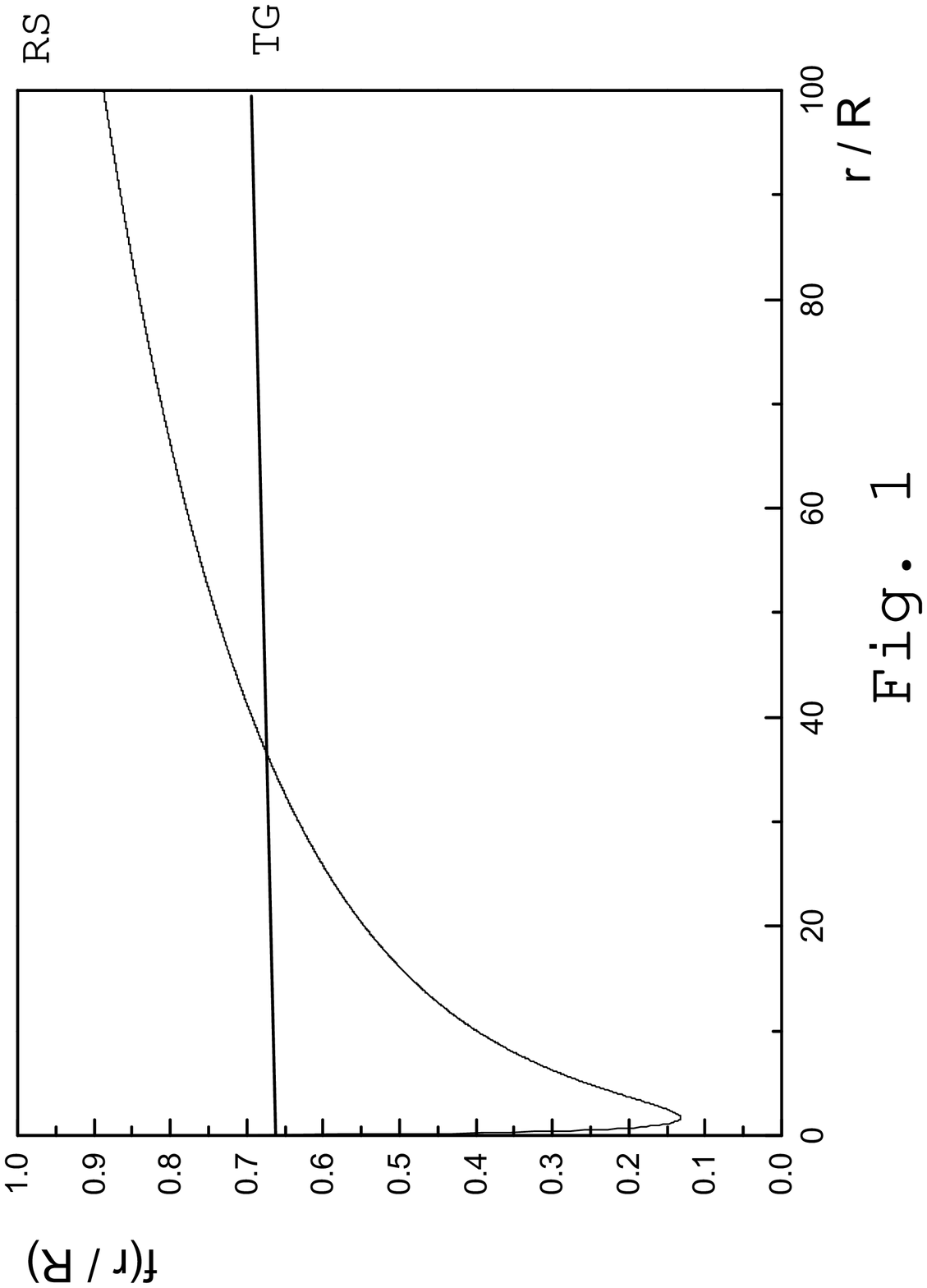}
%\newpage
%\epsfysize=20cm \epsfbox{Fig2.ps}


\begin{thebibliography}{99}

\bibitem{rs99-1} L. Randall and R. Sundrum, {\it Large mass hierarchy
from a extra dimension}, Phys. Rev. Lett. {\bf 83} (1999) 3370 [hep-th/9905221].

\bibitem{rs99-2} L. Randall and R. Sundrum, {\it An alternative to 
compactification}, 
Phys. Rev. Lett. {\bf B 83} (1999) 4690 [hep-th/9906064].

\bibitem{garr00} J. Garriga and T. Tanaka, {\it Gravity in the Randall-Sundrum
Brane World}, Phys. Rev. Lett. {\bf 84} (2000) 2778 [hep-th/9911055].

\bibitem{gidd00} S. B. Giddings, E. Katz, and L. Randall, {\it Linearized 
Gravity
in Brane Backgrounds}, JHEP {\bf 0003} (2000) 023 [hep-th/0002091].

\bibitem{duff00} M. J. Duff and J. T. Liu, {\it Complementarity of the 
Maldacena and
Randall-Sundrum Pictures}, Phys. Rev. Lett. {\bf 85} (2000) 2052 
[hep-th/0003237].

\bibitem{park01} D. K. Park and S. Tamaryan, {\it Compromise of Localized 
Graviton
with a Small Cosmological Constant in Randall-Sundrum Scenario}, 
Phys. Lett. {\bf B532} (2002) 305 [hep-th/0108068].

\bibitem{park02} D. K. Park and H. S. Kim, {\it Singular quantum mechanical
viewpoint of localized gravity in brane-world scenario}, Nucl. Phys. 
{\bf B 636} (2002) 179 [hep-th/0204122].

\bibitem{csa00} C. Cs\'{a}ki, J. Erlich, T. J. Hollowood and Y. Shirman, 
{\it Universal Aspects of Gravity Localized on Thick Branes}, Nucl. Phys.
{\bf B 581} (2000) 309 [hep-th/0001033].

\bibitem{fey65} R. P. Feynman and A. R. Hibbs, {\it Quantum Mechanics and 
Path Integrals}, McGraw-Hill, New York (1965).




\end{thebibliography}
\end{document}